\documentclass[preprint,times,12pt]{elsarticle}
\graphicspath{{simulation_graph/}}

\usepackage{graphicx}
\usepackage{amssymb}
\usepackage{txfonts}
\usepackage{pifont}
\usepackage{fleqn}
\usepackage{natbib}
\usepackage{geometry}
\usepackage{stmaryrd}

\newtheorem{thm}{Theorem}
\newtheorem{col}{Corollary}

\newdefinition{rmk}{Remark}
\newdefinition{exm}{Example}
\newtheorem{df}{Definition}
\newproof{pf}{\bf{Proof}}

\biboptions{numbers,square,sort&compress}

\journal{Information Sciences}

\begin{document}

\begin{frontmatter}

\title{Adaptive guaranteed-performance consensus design for high-order multiagent systems\tnoteref{t1}}
\tnotetext[t1]{This work was supported by the National Natural
Science Foundation of China under Grants 61374054, 61503012, 61703411, 61503009, 61333011 and 61421063 and Innovation Foundation of High-Tech Institute of Xi'an (2015ZZDJJ03), also supported by Innovation Zone Project (17-163-11-ZT-004-017-01).}

\author[a]{Jianxiang~Xi}
\author[a]{Jie~Yang}
\author[b]{Hao Liu\corref{cor1}}
\ead{liuhao13@buaa.edu.cn}
\author[a]{Tang Zheng}

\cortext[cor1]{Corresponding author. Tel.: +86 10 62790793.}
\address[a]{High-Tech Institute of Xi'an, Xi'an, 710025, P.R. China }
\address[b]{School of Astronautics, Beihang University, Beijing, 100191, P.R. China\vspace{-24pt}}

\begin{abstract}
The current paper addresses the distributed guaranteed-performance consensus design problems for general high-order linear multiagent systems with leaderless and leader-follower structures, respectively. The information about the Laplacian matrix of the interaction topology or its minimum nonzero eigenvalue is usually required in existing works on the guaranteed-performance consensus, which means that their conclusions are not completely distributed. A new translation-adaptive strategy is proposed to realize the completely distributed guaranteed-performance consensus control by using the structure feature of a complete graph in the current paper. For the leaderless case, an adaptive guaranteed-performance consensualization criterion is given in terms of Riccati inequalities and a regulation approach of the consensus control gain is presented by linear matrix inequalities. Extensions to the leader-follower cases are further investigated. Especially, the guaranteed-performance costs for leaderless and leader-follower cases are determined, respectively, which are associated with the intrinsic structure characteristic of the interaction topologies. Finally, two numerical examples are provided to demonstrate theoretical results.
\end{abstract}


\end{frontmatter}

\section{Introduction}\label{section1}
In the last two decades, the distributed cooperative control of multiagent systems has received great attention by researchers from different fields, such as formation control, multiple mobile robot systems, distributed computation and distributed sensor networks \cite{c1}-\cite{c9}, {\it et al}. Consensus is a fundamental problem in cooperative control, which requires that all agents achieve an agreement on certain variables of interest by designing distributed consensus protocols. Generally speaking, according to the structures of interaction topologies, consensus can be categorized into the leaderless one and the leader-follower one. When leader-follower consensus is achieved, the states of all followers track the state trajectories of the leader. However, for the leaderless case, all agents jointly determine the agreement state via interacting with each other, which is also called the consensus function. Some very interesting and significant works were finished in \cite{c10}-\cite{c18}, where the consensus regulation performance was not considered.\par

For many practical multiagent systems, it is not only required that they can achieve consensus, but also satisfy certain consensus performance. Hu {\it et al}. \cite{c19} showed a typical example; that is, when multiple mobile autonomous vehicles perform a specific patrol task, the distance performance is critically important due to limited resource or utility maximization. The consensus performance can be modeled as certain cost functions, which can be usually divided into two types: the individual cost function and the global cost function. For the individual cost function shown in \cite{c20} and \cite{c22}, each agent has a local objective function and some global goals can be achieved by optimizing the objective function of each agent. For the global cost function, some whole performance index is minimized in a distributed manner. For first-order multiagent systems, Cao and Ren \cite{c23} proposed a linear quadratic global cost function to achieve optimal consensus. For second-order multiagent systems, a global cost function was constructed by state errors among neighboring agents in \cite{c24}. For high-order multiagent systems, guaranteed-cost consensus criteria were presented in \cite{c25,c26,c27,c29}, which intrinsically realize suboptimal consensus control. It should be pointed out that the above consensus results with certain cost functions are not completely distributed since they are associated with the Laplacian matrix of the interaction topology or its nonzero eigenvalues, which are global information of a multiagent system as a whole.\par

For high-order multiagent systems without consensus performance constraints, Li {\it et al.} \cite{c30} proposed a very interesting dilation-adaptive strategy to eliminate the impacts of the nonzero eigenvalue of the Laplacian matrix and showed that the interaction strengths among agents may be monotonically increasing and converge to some finite steady-state value, but the quantitative value cannot be determined. Since the dilation factor is inversely proportional to the minimum nonzero eigenvalue, it may be very large when the algebraic connectivity is small and cannot be precisely determined. However, the precise value of the dilation factor is required to optimize the global cost function. Hence, the dilation-adaptive strategy is no longer valid to deal with guaranteed-performance consensus. To the best of our knowledge, the following three challenging problems on guaranteed-performance consensus are still open: (i) How to realize guaranteed-performance consensus control in a completely distributed manner; (ii) How to regulate the consensus control gain and consensus performance between any two agents; (iii) How to determine the guaranteed-performance cost when the interaction strengths among agents adaptively change and may be moronically increasing.\par

In the current paper, we intend to propose a distributed guaranteed-performance consensus scheme without using the Laplacian matrix and its minimum nonzero eigenvalue. Firstly, the new adaptive guaranteed-performance consensus protocols for leaderless cases and leader-follower cases are constructed, respectively, where interaction weights among neighboring agents are adaptively regulated by the state error between each agent and its neighbors for the leaderless case, but interaction weights between the leader and followers are time-varying and interaction weights among followers are fixed for the leader-follower case. Then, for leaderless and leader-follower cases, adaptive guaranteed-performance consensualization criteria are given, respectively, and the upper bounds of the guaranteed-performance costs are determined, respectively, which are independent of the adaptively regulated interaction weights. Finally, by combining with the linear matrix inequality (LMI) techniques, the regulation approaches of the consensus control gain are respectively given for leaderless and leader-follower multiagent systems by adjusting the translation factors.\par

Compared with closely related literatures on guaranteed-performance consensus for high-order linear multiagent systems, the current paper has three unique features. Firstly, the current paper proposes a novel translation-adaptive strategy to realize completely distributed guaranteed-performance consensus. However, the guaranteed-performance consensus criteria in \cite{c25,c26,c27,c29} require the global information and the dilation-adaptive strategy in \cite{c30} cannot be used to deal with guaranteed-performance constraints. Secondly, by the intrinsic structure property of the interaction topology, the regulation approach of the consensus control gain is proposed and the guaranteed-performance cost is determined in the current paper, while the consensus control gain cannot be regulated in \cite{c25,c26,c27,c29}, where the approaches to determine the guaranteed-performance cost are no longer valid when interaction weights are adaptively regulated. Thirdly, the consensus protocol in the current paper regulates the consensus performance between any two agents except the leader, but the consensus protocol in \cite{c25,c26,c27,c29} can only ensure the consensus performance between neighboring agents.\par

The remainder of the current paper is organized as follows. Section \ref{section2} presents leaderless adaptive guaranteed-performance consensualization criteria, determines the guaranteed-performance cost, and gives a consensus control gain regulation approach. In Section \ref{section3}, leader-follower adaptive guaranteed-performance consensus problems are investigated and the associated consensualization criteria are proposed. Two numerical examples are given to demonstrate theoretical results in Section \ref{section4} and some concluding remarks are presented in Section \ref{section5}.\par

{\it Notations}: ${\mathbb{R}^d}$ denotes the real column vector space of dimension $d$ and ${\mathbb{R}^{d \times d}}$ stands for the set of $d \times d$ dimensional real matrices. ${I_N}$ represents the identity matrix of dimension $N$ and ${{\bf{1}}_N}$ is an $N$-dimensional column vector with all entries equal to 1. ${\bf{0}}$ stands for the zero column vector with a compatible dimension. ${Q^T}$ denotes the transpose of $Q$. ${R^T} = R > {\rm{0}}$ and ${R^T} = R < {\rm{0}}$ mean that the symmetric matrix $R$ is positive definite and negative definite, respectively. ${R^T} = R \ge {\rm{0}}$ and ${R^T} = R \le {\rm{0}}$ mean that the symmetric matrix $R$ is positive semidefinite and negative semidefinite, respectively. The notation $ \otimes $ stands for the Kronecker product. The symmetric elements of a symmetric matrix are denoted by the symbol *.\par

\section{Leaderless adaptive guaranteed-performance consensualization}\label{section2}
This section first presents the problem description for leaderless adaptive guaranteed-performance consensus design. Then, the associated consensualization criteria are given in terms of Ricatti inequalities, the guaranteed-performance cost is determined and an approach to regulate the consensus control gain is shown in terms of LMIs.

\subsection{Problem description for leaderless cases}
A connected undirected graph $G$ is used to describe the interaction topology of a multiagent system, where each agent is denoted by a node, the interaction channel between any two nodes is represented by an edge and the interaction strength is denoted the edge weight ${l_{ik}}{w_{ik}}(t)$, where ${l_{ii}} = 0$, ${l_{ik}} = 1$ if agent $k$ is a neighbor of agent $i$ and ${l_{ik}} \equiv 0$ otherwise, and ${w_{ik}}(t) > 0$ with ${w_{ik}}(0) > 0$ is a nondecreasing function designed later. The Laplacian matrix of $G$ is defined as ${L_{w(t)}} = D(t) - W(t)$, where $W(t) = {\left[ {{l_{ik}}{w_{ik}}(t)} \right]_{N \times N}}$ denotes the weight matrix of $G$ and $D(t) = {\rm{diag}}\left\{ {{d_1}(t),{d_2}(t), \cdots ,{d_N}(t)} \right\}$ with ${d_i}(t) = \sum\nolimits_{k = 1,k \ne i}^N {{l_{ik}}{w_{ik}}(t)} \left( {i = 1,2, \cdots ,N} \right)$ represents the in-degree matrix of $G$. Since $G$ is connected, zero is its single eigenvalue and the other eigenvalues are positive as shown in \cite{c31}.\par

The dynamics of each agent is described by the following general high-order linear model:
\begin{eqnarray}\label{1}
{\dot x_i}(t) = A{x_i}(t) + B{u_i}(t){\rm{ }}\left( {i \in \left\{ {1,2, \cdots ,N} \right\}} \right),
\end{eqnarray}
where $A \in {\mathbb{R}^{d \times d}},$ $B \in {\mathbb{R}^{d \times p}},$ $N$ is a positive integer standing for the number of agents, and ${x_i}(t)$ and ${u_i}(t)$ are the state and the control input of agent $i$, respectively.\par

A leaderless adaptive guaranteed-performance consensus protocol is proposed as follows
\vspace{1em}
\begin{eqnarray}\label{2}
\left\{ \begin{array}{l}
{u_i}(t) = {K_u}\sum\limits_{k \in {N_i}} {{w_{ik}}(t)\left( {{x_k}(t) - {x_i}(t)} \right)} ,\\
{{\dot w}_{ik}}(t) = {\left( {{x_k}(t) - {x_i}(t)} \right)^T}{K_w}\left( {{x_k}(t) - {x_i}(t)} \right),\\
{J_r} = \frac{1}{N}\sum\limits_{i = 1}^N {\sum\limits_{k = 1}^N {\int_0^{ + \infty } {{{\left( {{x_k}(t) - {x_i}(t)} \right)}^T}Q\left( {{x_k}(t) - {x_i}(t)} \right){\rm{d}}t} } } ,
\end{array} \right.
\end{eqnarray}
where ${K_u} \in {\mathbb{R}^{p \times d}}$ and ${K_w} \in {\mathbb{R}^{d \times d}}$ are gain matrices with $K_w^T = {K_w} \ge 0$, ${N_i}$ denotes the neighbor set of agent $i$ and ${Q^T} = Q > 0$. Here, it is assumed that there exists an upper bound ${\gamma _{ik}}$ of ${w_{ik}}(t)$~$(i,k \in \{ 1,2, \cdots ,N\} )$, which can be determined by (\ref{2}) and is related to the state error between two neighboring agents and the unknown gain matrix ${K_w}$. Protocol (\ref{2}) has two new features. The first one is that ${w_{ik}}(t)$ with ${w_{ik}}(0) >0$ is adaptively regulated and is nondecreasing since ${K_w}$ is symmetric and positive semidefinite. If agent $k$ is a neighbor of agent $i$, then ${w_{ik}}(t)$ is a practical interaction weight from agent $k$ to agent $i$. Otherwise, ${w_{ik}}(t)$ can be regarded as a virtual interaction weight from agent $k$ to agent $i$. The second one is that the consensus performance of all agents instead of neighboring agents can be regulated.\par

The definition of the leaderless adaptive guaranteed-performance consensualization is given as follows.
\begin{df}\label{definition1}
\textup {Multiagent system (\ref{1}) is said to be leaderless adaptively guaranteed-perfor-mance consensualizable by protocol (\ref{2}) if there exist ${K_u}$ and ${K_w}$ such that ${\lim _{t \to  + \infty }}({x_i}(t) - $ ${x_k}(t))  \hspace{-2pt}= \hspace{-2pt} {\bf{0}}{\rm{ }}\left( {i,k \hspace{-2pt} =  \hspace{-2pt} 1,2, \cdots ,N} \right)$ and ${J_r} \le J_r^*$ for any bounded initial states ${x_i}(0){\rm{ }}\left( {i \hspace{-2pt} =  \hspace{-2pt}1,2, \cdots ,N} \right)$, where $J_r^*$ is said to be the guaranteed-performance cost.}
\end{df}

\begin{rmk}\label{remark0}
In protocol (2), ${J_r}(t)$ denotes the consensus performance regulation term, which can be realized by choosing a proper matrix $Q$. The $j$th diagonal element of $Q$ stands for the optimization weight of the $j$th component of the state error between agents $k$ and $i$ and the other elements of $Q$ represent the coupling relationships among the corresponding components of the state error. Especially, for practical multiagent systems, the matrix $Q$ is usually chosen as a diagonal matrix. In this case, a bigger optimization weight can guarantee the smaller squared sum of the associated component of the state error by the controller design approaches, which are often based on the Riccati inequality. More detailed explanations and theoretical analysis about the impacts of the parameters to the control performance for isolated systems and multiagent systems can be found in \cite{c23} and \cite{c33}, respectively.
\end{rmk}

In the following, we design gain matrices ${K_u}$ and ${K_w}$ such that multiagent system (\ref{1}) achieves leaderless adaptive guaranteed-performance consensus, and determine the guaranteed-performance cost $J_r^*$. Furthermore, an approach to regulate the consensus control gain is proposed and the consensus motion is determined.

\subsection{Adaptive guaranteed-performance consensus design for leaderless cases}
Let $x(t) = {\left[ {x_1^T(t),x_2^T(t), \cdots ,x_N^T(t)} \right]^T}{\rm{ ,}}$ then the dynamics of multiagent system (\ref{1}) with protocol (\ref{2}) can be written in a compact form as
\begin{eqnarray}\label{3}
\dot x(t) = \left( {{I_N} \otimes A - {L_{w(t)}} \otimes B{K_u}} \right)x(t),
\end{eqnarray}
where ${L_{w(t)}}$ is the Laplacian matrix of the interaction topology. Since the interaction topology is connected, there exists an orthonormal matrix $U = \left[ {{{{{\bf{1}}_N}} \mathord{\left/
 {\vphantom {{{{\bf{1}}_N}} {\sqrt N }}} \right.
 \kern-\nulldelimiterspace} {\sqrt N }},\bar U} \right]$ such that
\begin{eqnarray}\label{4}
{U^T}{L_{w(0)}}U = \left[ {\begin{array}{*{20}{c}}
0&{{{\bf{0}}^T}}\\
{\bf{0}}&{{\Delta _{w(0)}}}
\end{array}} \right],
\end{eqnarray}
where ${\Delta _{w(0)}} = {\bar U^T}{L_{w(0)}}\bar U = {\rm{diag}}\left\{ {{\lambda _2},{\lambda _3}, \cdots ,{\lambda _N}} \right\}$ with ${\lambda _2} \le {\lambda _3} \le  \cdots  \le {\lambda _N}$ being nonzero eigenvalues of ${L_{w(0)}}$. Let $\bar x(t) = \left( {{U^T} \otimes {I_d}} \right)x(t) = {\left[ {\bar x_1^T(t),{\eta ^T}(t)} \right]^T}$ with $\eta (t) = [\bar x_2^T(t),\bar x_3^T(t), \cdots ,$ $\bar x_N^T(t){]^T}$, then multiagent system (3) can be transformed into
\begin{eqnarray}\label{5}
{\dot {\bar x}_1}(t) = A{\bar x_1}(t),
\end{eqnarray}
\begin{eqnarray}\label{6}
\dot {\eta} (t) = \left( {{I_{N - 1}} \otimes A - {{\bar U}^T}{L_{w(t)}}\bar U \otimes B{K_u}} \right)\eta (t).
\end{eqnarray}
Let ${e_i}$ $(i = 1,2, \cdots ,N)$ denote $N$ -dimensional column vectors with the $i$th element 1 and 0 elsewhere, then we define
\begin{eqnarray}\label{7}
{x_{\bar c}}(t) \buildrel \Delta \over = \sum\limits_{i = 2}^N {U{e_i} \otimes {{\bar x}_i}(t)},
\end{eqnarray}
\begin{eqnarray}\label{8}
{x_c}(t) \buildrel \Delta \over = U{e_1} \otimes {\bar x_1}(t) = \frac{1}{{\sqrt N }}{{\bf{1}}_N} \otimes {\bar x_1}(t).
\end{eqnarray}
Due to
\[\sum\limits_{i = 2}^N {{e_i} \otimes {{\bar x}_i}(t)}  = {\left[ {{{\bf{0}}^T},{\eta ^T}(t)} \right]^T},\]
it can be shown by (\ref{7}) that
\begin{eqnarray}\label{9}
{x_{\bar c}}(t) = \left( {U \otimes {I_d}} \right){\left[ {{{\bf{0}}^T},{\eta ^T}(t)} \right]^T}.
\end{eqnarray}
From (\ref{8}), one can see that
\begin{eqnarray}\label{10}
{x_c}(t) = \left( {U \otimes {I_d}} \right){\left[ {\bar x_1^T(t),{{\bf{0}}^T}} \right]^T}.
\end{eqnarray}
Because $U \otimes {I_d}$ is nonsingular, ${x_{\bar c}}(t)$ and ${x_c}(t)$ are linearly independent by (\ref{9}) and (\ref{10}). Due to $\left( {{U^T} \otimes {I_d}} \right)x(t) = {\left[ {\bar x_1^T(t),{\eta ^T}(t)} \right]^T}$, one can obtain that $x(t) = {x_{\bar c}}(t) + {x_c}(t).$ By the structure of ${x_c}(t)$ in (\ref{8}), multiagent system (\ref{1}) achieves consensus if and only if ${\lim _{t \to  + \infty }}\eta (t) = {\bf{0}}$; that is, subsystems (\ref{5}) and (\ref{6}) describe the consensus motion and relative state motion of multiagent system (\ref{1}), respectively.\par

The following theorem presents a sufficient condition for leaderless adaptive guaranteed-performance consensualization, which can realize completely distributed guaranteed-per-formance consensus design; that is, the design approach of gain matrices in protocol (\ref{2}) is independent of the time-varying interaction topology and its eigenvalues.

\begin{thm}\label{theorem1}
For any given translation factor $\gamma  > 0$, multiagent system (\ref{1}) is adaptively guaranteed-performance consensualizable by protocol (\ref{2}) if there exists a matrix ${P^T} = P > 0$ such that $PA + {A^T}P - \gamma PB{B^T}P + 2Q \le 0$. In this case, ${K_u} = {B^T}P$, ${K_w} = PB{B^T}P$ and the guaranteed-performance cost satisfies that
\[
J_r^* = {x^T}(0)\left( {\left( {{I_N} - \frac{1}{N}{{\bf{1}}_N}{\bf{1}}_N^T} \right) \otimes P} \right)x(0) + \gamma \int_0^{ + \infty } {{x^T}(t)\left( {\left( {{I_N} - \frac{1}{N}{{\bf{1}}_N}{\bf{1}}_N^T} \right) \otimes PB{B^T}P} \right)x(t){\rm{d}}t.}
\]
\end{thm}

\begin{pf}
First of all, we design ${K_u}$ and ${K_w}$ such that ${\lim _{t \to  + \infty }}\eta (t) = {\bf{0}}$. Construct the following new Lyapunov function candidate
\[
V(t) = {\eta ^T}(t)\left( {{I_{N - 1}} \otimes P} \right)\eta (t) + \sum\limits_{i = 1}^N {\sum\limits_{k \in {N_i}} {\frac{{{{\left( {{w_{ik}}(t) - {w_{ik}}(0)} \right)}^2}}}{2}} }  + \frac{\gamma }{{2N}}\sum\limits_{i = 1}^N {\sum\limits_{k = 1,k \ne i}^N {\left( {{\gamma _{ik}} - {w_{ik}}(t)} \right)} } .
\]
Due to ${P^T} = P > 0$ and ${\gamma _{ik}} \ge {w_{ik}}(t)$, one has $V(t) \ge 0$. Since it is assumed that the interaction topology is undirected, one has ${L_{w(t)}} = L_{w(t)}^T$ . Let ${K_u} = {B^T}P$, then the time derivative of $V(t)$ along the solution of subsystem (\ref{6}) is
\[
\dot V(t) = {\eta ^T}(t)\left( {{I_{N - 1}} \otimes \left( {PA + {A^T}P} \right) - 2{{\bar U}^T}{L_{w(t)}}\bar U \otimes PB{B^T}P} \right)\eta (t)
\]\vspace{-1.5em}
\begin{eqnarray}\label{11}
~~~~~~~~~~+ \sum\limits_{i = 1}^N {\sum\limits_{k \in {N_i}} {\left( {{w_{ik}}(t) - {w_{ik}}(0)} \right)} {{\dot w}_{ik}}(t)}  - \frac{\gamma }{{2N}}\sum\limits_{i = 1}^N {\sum\limits_{k = 1,k \ne i}^N {{{\dot w}_{ik}}(t)} }.
\end{eqnarray}
Due to $U{U^T} = {I_N}$, one can show that $\bar U{\bar U^T} = {L_N}$, where ${L_N}$ is the Laplacian matrix of a complete graph with the weights of all the edges ${1 \mathord{\left/
 {\vphantom {1 N}} \right.
 \kern-\nulldelimiterspace} N}$. Thus, it can be obtained by (\ref{2}) and (\ref{4}) that
 \[
~~ {\rm{  }}\sum\limits_{i = 1}^N {\sum\limits_{k \in {N_i}} {\left( {{w_{ik}}(t) - {w_{ik}}(0)} \right)} {{\dot w}_{ik}}(t)}  - \frac{\gamma }{{2N}}\sum\limits_{i = 1}^N {\sum\limits_{k = 1,k \ne i}^N {{{\dot w}_{ik}}(t)} }
 \]\vspace{-1.25em}
 \[
  = {x^T}(t)\left( {\left( {2{L_{w(t)}} - 2{L_{w(0)}} - \gamma {L_N}} \right) \otimes {K_w}} \right)x(t)
  \]\vspace{-1.25em}
\begin{eqnarray}\label{12}
 = {\eta ^T}(t)\left( {\left( {2{{\bar U}^T}{L_{w(t)}}\bar U - 2{\Delta _{w(0)}} - \gamma {I_{N - 1}}} \right) \otimes {K_w}} \right)\eta (t).
\end{eqnarray}
Let ${K_w} = PB{B^T}P$, then one can derive from (\ref{11}) and (\ref{12}) that
\[\dot V(t) \le \sum\limits_{i = 2}^N {\bar x_i^T(t)\left( {PA + {A^T}P - \left( {2{\lambda _i} + \gamma } \right)PB{B^T}P} \right){{\bar x}_i}(t)} .\]
Due to $\gamma  > 0$ and ${\lambda _i} > 0{\rm{ }}\left( {i = 2,3, \cdots ,N} \right)$, one can obtain that if $PA + {A^T}P - \gamma PB{B^T}P < 0$, then
\[\dot V(t) \le  - \varepsilon \sum\limits_{i = 2}^N {\bar x_i^T(t){{\bar x}_i}(t)}  =  - \varepsilon {\left\| {\eta (t)} \right\|^2}\]
for some positive real constant $\varepsilon $. Therefore, $\eta (t)$ converges to ${\bf{0}}$ asymptotically, which means that multiagent system (\ref{1}) achieves leaderless adaptive consensus.\par
In the following, we determine the guaranteed-performance cost. It can be shown that
\begin{eqnarray}\label{13}
\sum\limits_{i = 1}^N {\sum\limits_{k = 1}^N {{{\left( {{x_k}(t) - {x_i}(t)} \right)}^T}Q\left( {{x_k}(t) - {x_i}(t)} \right)} }  = {x^T}(t)\left( {2N{L_N} \otimes Q} \right)x(t).
\end{eqnarray}
Due to ${x^T}(t)\left( {{L_N} \otimes I} \right)x(t) = {\eta ^T}(t)\eta (t),$ it can be derived that
\begin{eqnarray}\label{14}
{x^T}(t)\left( {{L_N} \otimes Q} \right)x(t) \le \sum\limits_{i = 2}^N {\bar x_i^T(t)Q{{\bar x}_i}(t).}
\end{eqnarray}
Let $h > 0$, then one can obtain by (\ref{13}) and (\ref{14}) that
\begin{eqnarray}\label{15}
J_r^h \buildrel \Delta \over = \frac{1}{N}\sum\limits_{i = 1}^N {\sum\limits_{k = 1}^N {\int_0^h {{{\left( {{x_k}(t) - {x_i}(t)} \right)}^T}Q\left( {{x_k}(t) - {x_i}(t)} \right){\rm{d}}t} } }  \le \sum\limits_{i = 2}^N {\int_0^h {2\bar x_i^T(t)Q{{\bar x}_i}(t){\rm{d}}t} } .
\end{eqnarray}
Moreover, one can show that
\begin{eqnarray}\label{16}
\int_0^h {\dot V(t)} {\rm{d}}t - V(h) + V(0) = 0.
\end{eqnarray}
Due to ${\lim _{t \to  + \infty }}\left( {{w_{ik}}(t) - {\gamma _{ik}}} \right) = 0$, one has
\begin{eqnarray}\label{17}
\mathop {\lim }\limits_{h \to  + \infty } \sum\limits_{i = 1}^N {\sum\limits_{k = 1,k \ne i}^N {\left( {{\gamma _{ik}} - {w_{ik}}(h)} \right)} }  = 0.
\end{eqnarray}
If $PA + {A^T}P - \gamma PB{B^T}P + 2Q \le 0$, then one can obtain from (\ref{15}) to (\ref{17}) that
\begin{eqnarray}\label{18}
\mathop {\lim }\limits_{h \to  + \infty } J_r^h \le {\eta ^T}(0)\left( {{I_{N - 1}} \otimes P} \right)\eta (0) + \frac{\gamma }{{2N}}\sum\limits_{i = 1}^N {\sum\limits_{k = 1,k \ne i}^N {\left( {{\gamma _{ik}} - {w_{ik}}(0)} \right)} } .
\end{eqnarray}
Since $\eta (t) = \left[ {{{\bf{0}}_{(N - 1)d \times d}},{I_{(N - 1)d}}} \right]\left( {{U^T} \otimes {I_d}} \right)x(t)$ and $\bar U{\bar U^T} = {L_N}$, one has
\begin{eqnarray}\label{19}
{\eta ^T}(0)\left( {{I_{N - 1}} \otimes P} \right)\eta (0) = {x^T}(0)\left( {\left( {{I_N} - \frac{1}{N}{{\bf{1}}_N}{\bf{1}}_N^T} \right) \otimes P} \right)x(0).
\end{eqnarray}
Due to ${\lim _{t \to  + \infty }}\left( {{w_{ik}}(t) - {\gamma _{ik}}} \right) = 0$, one can show that
\begin{eqnarray}\label{20}
\sum\limits_{i = 1}^N {\sum\limits_{k = 1,k \ne i}^N {\left( {{\gamma _{ik}} - {w_{ik}}(0)} \right)} }  = \sum\limits_{i = 1}^N {\sum\limits_{k = 1,k \ne i}^N {\int_0^{ + \infty } {{{\dot w}_{ik}}(t){\rm{d}}t} } }  = 2N\int_0^{ + \infty } {{x^T}(t)\left( {{L_N} \otimes {K_w}} \right)x(t){\rm{d}}t} .
\end{eqnarray}
From (\ref{18}) to (\ref{20}), the conclusion of Theorem \ref{theorem1} can be obtained.$\Box$
\end{pf}

In \cite{c33}, it was shown that the Riccati equation $PA + {A^T}P - \gamma PB{B^T}P + 2Q = 0$ has a unique and positive definite solution $P$ for any given $\gamma  > 0$ if $(A,B)$ is stabilizable. Moreover, from the proof of Theorem \ref{theorem1}, we eliminate the impacts of the nonzero eigenvalues of ${L_{w(0)}}$ by introducing a positive constant $\gamma $ to construct the term $\left( {2{\lambda _i} + \gamma } \right)PB{B^T}P$ $\left( {i = 2,3, \cdots ,N} \right)$. Thus, $\gamma $ can be regarded as the rightward translated quantity of the nonzero eigenvalues of $2{L_{w(0)}}$ and can be given previously.\par

Furthermore, a large $\gamma $ may regulate the consensus control gain by Theorem \ref{theorem1}, so we can choose some proper $\gamma $ and $P$ to regulate the consensus control gain. We introduce a gain factor $\delta  > 0$ such that $P \le \delta I$, where $\delta $ can also be regarded as an upper bound of the eigenvalue of $P$. Thus, one can show that $PB{B^T}P \le {\delta ^2}B{B^T}$ if the maximum eigenvalue of $B{B^T}$ is not larger than 1. Based on LMI techniques, by Schur complement lemma in \cite{c32}, an adaptive guaranteed-performance consensualization criterion with a given gain factor is proposed as follows.\par

\begin{col}\label{corollary1}
For any given gain factor $\delta  > 0$, multiagent system (\ref{1}) is leaderless adaptively guaranteed-performance consensualizable by protocol (\ref{2}) if ${\lambda _{\max }}\left( {B{B^T}} \right) \le 1$ and there exist $\gamma  > 0$ and ${\tilde P^T} = \tilde P \ge {\delta ^{ - 1}}I$ such that
\[\tilde \Xi  = \left[ {\begin{array}{*{20}{c}}
{A\tilde P + \tilde P{A^T} - \gamma B{B^T}}&{2\tilde PQ}\\
*&{ - 2Q}
\end{array}} \right] < 0.\]
In this case, ${K_u} = {B^T}{\tilde P^{ - 1}}$, ${K_w} = {\tilde P^{ - 1}}B{B^T}{\tilde P^{ - 1}}$ and the guaranteed-performance cost satisfies that
\[J_r^* = \sum\limits_{i = 2}^N {\left( {\delta {{\left\| {{{\bar x}_i}(0)} \right\|}^2} + \gamma {\delta ^2}\int_0^{ + \infty } {{{\left\| {{B^T}{{\bar x}_i}(t)} \right\|}^2}} {\rm{d}}t} \right)} .\]
\end{col}

In the following, we give an approach to determine the consensus motion. Due to $e_1^T{U^T} = {{{\bf{1}}_N^T} \mathord{\left/
 {\vphantom {{{\bf{1}}_N^T} {\sqrt N }}} \right.
 \kern-\nulldelimiterspace} {\sqrt N }}$, one has\vspace{-0.25em}
 \[{\bar x_1}(0) = \left( {e_1^T \otimes {I_d}} \right)\left( {{U^T} \otimes {I_d}} \right)x(0) = \frac{1}{{\sqrt N }}\sum\limits_{i = 1}^N {{x_i}(0)}. \]\vspace{-0.25em}
 Because subsystem (\ref{5}) describes the consensus motion of multiagent system (\ref{1}), the following corollary can be obtained by (\ref{8}).\vspace{-0.5em}

\begin{col}\label{corollary2}
If multiagent system (\ref{1}) achieves leaderless adaptive guaranteed-performance consensus, then\vspace{-0.25em}
\[\mathop {\lim }\limits_{t \to  + \infty } \left( {{x_i}(t) - {e^{At}}\left( {\frac{1}{N}\sum\limits_{i = 1}^N {{x_i}(0)} } \right)} \right) = 0{\rm{ }}\left( {i = 1,2, \cdots ,N} \right).\]\vspace{-1.25em}
\end{col}

\begin{rmk}\label{remark1}
In \cite{c25,c26,c27,c29}, guaranteed-performance consensus criteria are associated with the Laplacian matrix of the interaction topology or its minimum and maximum nonzero eigenvalues, which means that global information of the whole system is required essentially and their approaches are not completely distributed. A novel dilation adaptive consensus strategy was proposed to realize completely distributed consensus control in \cite{c30} , where guaranteed-performance constraints were not considered and the dilation factor is inversely proportional to the minimum nonzero eigenvalue. When guaranteed-performance constraints are considered, the precise value of the dilation factor is required. However, to obtain the precise value of the dilation factor, it is necessary to determine the minimum nonzero eigenvalue. In Theorem \ref{theorem1}, the translation-adaptive strategy is proposed to deal with guaranteed-performance constraints without using information of the Laplacian matrix. Actually, we realize the eigenvalue translation by the special property of the Laplacian matrix of a complete graph; that is, all its eigenvalues are identical.\vspace{-1em}
\end{rmk}

\section{Leader-follower adaptive guaranteed-performance consensualization}\label{section3}\vspace{-0.5em}
This section focuses on leader-follower adaptive guaranteed-performance consensus, where multiagent systems consist of a leader and $N - 1$ followers. Sufficient conditions for leader-follower adaptive guaranteed-performance consensualization are presented on the basis of the Ricatti inequality and LMIs, respectively, and the guaranteed-performance cost is determined.\par

\subsection{Problem description for leader-follower cases}
Without loss of generality, we set that agent 1 is the leader and the other $N - 1$ agents are followers. Thus, the dynamics of multiagent systems with the leader-follower structure can be described by
\vspace{-1em}
\begin{eqnarray}\label{21}
\left\{ \begin{array}{l}
{{\dot x}_1}(t) = A{x_1}(t),\\
{{\dot x}_i}(t) = A{x_i}(t) + B{u_i}(t),
\end{array} \right.
\end{eqnarray}
where $i = 2,3, \cdots ,N,$ $A \in {\mathbb{R}^{d \times d}},$ $B \in {\mathbb{R}^{d \times p}},$ ${x_1}(t)$ is the state of the leader, and ${x_i}(t)$ and ${u_i}(t)$ are the state and the control input of the $i$th follower, respectively. The leader does not receive any information from followers and there at least exists an undirected path from the leader to each follower. Furthermore, it is assumed that the subgraph among followers is undirected but can be unconnected.\par

A new consensus protocol is proposed to realize adaptive guaranteed-performance tracking as follows
\begin{eqnarray}\label{22}
\left\{ \begin{array}{l}
{u_i}(t) = {l_{i1}}{w_{i1}}(t){K_u}\left( {{x_1}(t) - {x_i}(t)} \right) + {K_u}\sum\limits_{k \in {N_i},k \ne 1} {\left( {{x_k}(t) - {x_i}(t)} \right)} ,\\
{{\dot w}_{i1}}(t) = {\left( {{x_1}(t) - {x_i}(t)} \right)^T}{K_w}\left( {{x_1}(t) - {x_i}(t)} \right),\\
{J_l} = {J_{fl}} + {J_{ff}},
\end{array} \right.
\end{eqnarray}
where $i = 2,3, \cdots ,N,$ ${K_u} \in {\mathbb{R}^{p \times d}}$ and ${K_w} \in {\mathbb{R}^{d \times d}}$ are gain matrices with $K_w^T = {K_w} \ge 0$, ${N_i}$ denotes the neighbor set of agent $i$, ${l_{11}} = 0$, ${l_{i1}} = 1$ if agent $i$ can receive the information of the leader and ${l_{i1}} \equiv 0$ otherwise, ${l_{i1}}{w_{i1}}(t)$ with ${w_{i1}}(0) = 1$ is the adaptively regulated interaction strength from the leader to the $i$th follower, it is supposed that the upper bound of ${w_{i1}}(t)$ is ${\gamma _{i1}}$, and for ${Q^T} = Q > 0$,
\[{J_{fl}} = \sum\limits_{i = 2}^N {\int_0^{ + \infty } {{l_{i1}}{{\left( {{x_1}(t) - {x_i}(t)} \right)}^T}Q\left( {{x_1}(t) - {x_i}(t)} \right){\rm{d}}t} }, \]
\[{J_{ff}} = \frac{1}{{N - 1}}\sum\limits_{i = 2}^N {\sum\limits_{k = 2}^N {\int_0^{ + \infty } {{{\left( {{x_k}(t) - {x_i}(t)} \right)}^T}Q\left( {{x_k}(t) - {x_i}(t)} \right){\rm{d}}t} } } .\]
From protocol (\ref{22}), interaction strengths from the leader to followers are adaptively time-varying, but interaction strengths among followers are time-invariant and are equal to 1. Now, we give the definition of the leader-follower adaptive guaranteed-performance consensualization as follows.

\begin{df}\label{definition2}
\textup {Multiagent system (\ref{21}) is said to be leader-follower adaptively guaranteed-performance consensualizable by protocol (\ref{22}) if there exist ${K_u}$ and ${K_w}$ such that ${\lim _{t \to  + \infty }}$ $({x_i}(t) - {x_1}(t)) = {\bf{0}}{\rm{ }}\left( {i = 2,3, \cdots ,N} \right)$ and ${J_l} \le J_l^*$ for any bounded initial states ${x_i}(0){\rm{ (}}i = 1,2,$ $ \cdots ,N{\rm{)}}$, where $J_l^*$ is said to be the guaranteed-performance cost.}
\end{df}

In the following, an approach is given to design gain matrices ${K_u}$ and ${K_w}$ such that multiagent system (\ref{21}) with protocol (\ref{22}) achieves leader-follower adaptive guaranteed-performance consensus and the guaranteed-performance cost $J_l^*$ is determined. Furthermore, it is revealed that the consensus control gain can be regulated by introducing a gain factor.

\subsection{Adaptive guaranteed-performance consensus design for leader-follower cases}
Let ${\xi _i}(t) = {x_i}(t) - {x_1}(t){\rm{ }}\left( {i = 2,3, \cdots ,N} \right)$ and $\xi (t) = {\left[ {\xi _2^T(t),\xi _3^T(t), \cdots ,\xi _N^T(t)} \right]^T}$, then one can obtain by (\ref{21}) and (\ref{22}) that
\begin{eqnarray}\label{23}
\dot \xi (t) = \left( {{I_{N - 1}} \otimes A - \left( {{L_{ff}} + {\Lambda _{w(t),fl}}} \right) \otimes B{K_u}} \right)\xi (t),
\end{eqnarray}
where ${L_{ff}}$ is the Laplacian matrix of the interaction topology among followers and ${\Lambda _{w(t),fl}} = {\rm{diag}}\left\{ {{l_{21}}{w_{21}}(t),{l_{31}}{w_{31}}(t), \cdots ,{l_{N1}}{w_{N1}}(t)} \right\}$ denotes the interaction from the leader to followers. If ${\lim _{t \to  + \infty }}\xi (t) = {\bf{0}}$, then multiagent (\ref{21}) with protocol (\ref{22}) achieves leader-follower consensus. Let $x(t) = {\left[ {x_1^T(t),x_2^T(t), \cdots ,x_N^T(t)} \right]^T}$, then the following theorem presents a leader-follower adaptive guaranteed-performance consensualization criterion and determines the guaranteed-performance cost.

\begin{thm}\label{theorem2}
For any given translation factor ${\gamma _l} > 0$, multiagent system (\ref{21}) is leader-follower adaptively guaranteed-performance consensualizable by protocol (\ref{22}) if there exists a matrix ${R^T} = R > 0$ such that $RA + {A^T}R - {\gamma _l}RB{B^T}R + 3Q \le 0$. In this case, ${K_u} = {B^T}R$, ${K_w} = RB{B^T}R$ and the guaranteed-performance cost satisfies that
\[
J_l^* \hspace{-2pt}= \hspace{-2pt}{x^T}(0)\left( {\left[ {\begin{array}{*{20}{c}}
   {N - 1} & { - {\bf{1}}_{N - 1}^T}  \\
   { - {{\bf{1}}_{N - 1}}} & {{I_{N - 1}}}  \\
\end{array}} \right] \hspace{-2pt}\otimes \hspace{-2pt} R} \right)x(0) \hspace{-2pt}+\hspace{-2pt} {\gamma _l}\int_0^{ + \infty }\hspace{-4pt} {{x^T}(t)\left( {\left[ {\begin{array}{*{20}{c}}
   {N - 1} & { - {\bf{1}}_{N - 1}^T}  \\
   { - {{\bf{1}}_{N - 1}}} & {{I_{N - 1}}}  \\
\end{array}} \right]\hspace{-2pt} \otimes \hspace{-2pt} RB{B^T}R} \right)x(t){\rm{d}}t.}
\]
\end{thm}

\begin{pf}
Construct a new Lyapunov function candidate as follows
\[V(t) = {\xi ^T}(t)\left( {{I_{N - 1}} \otimes R} \right)\xi (t) + \sum\limits_{i = 2}^N {{l_{i1}}{{\left( {{w_{i1}}(t) - {w_{ik}}(0)} \right)}^2}}  + {\gamma _l}\sum\limits_{i = 2}^N {\left( {{\gamma _{i1}} - {w_{i1}}(t)} \right)} .\]
Due to ${R^T} = R > 0$ and ${\gamma _{i1}} \ge {w_{i1}}(t){\rm{ }}(i = 2,3, \cdots ,N)$, one can obtain that $V(t) \ge 0$. Let ${K_u} = {B^T}R$ and ${K_w} = RB{B^T}R$, then the derivative of $V(t)$ with respect to time $t$ can be given by (\ref{23}) as follows
\[
\dot V(t) = {\xi ^T}(t)\left( {{I_{N - 1}} \otimes \left( {RA + {A^T}R} \right) - 2\left( {{L_{ff}} + {\Lambda _{w(t),fl}}} \right) \otimes RB{B^T}R} \right)\xi (t)
\]\vspace{-1.25em}
\begin{eqnarray}\label{24}
~~~~~~~~~~~+ \sum\limits_{i = 2}^N {2{l_{i1}}\left( {{w_{i1}}(t) - {w_{i1}}(0)} \right)} {{\dot w}_{i1}}(t) - {\gamma _l}\sum\limits_{i = 2}^N {{{\dot w}_{i1}}(t)} .
\end{eqnarray}
By (\ref{22}), then one can show that
\begin{eqnarray}\label{25}
\sum\limits_{i = 2}^N {{{\dot w}_{i1}}(t)}  = {\xi ^T}(t)\left( {{I_{N - 1}} \otimes RB{B^T}R} \right)\xi (t),
\end{eqnarray}
\begin{eqnarray}\label{26}
\sum\limits_{i = 2}^N {{l_{i1}}\left( {{w_{i1}}(t) - {w_{i1}}(0)} \right)} {\dot w_{i1}}(t) = {\xi ^T}(t)\left( {{\Lambda _{w(t),fl}} - {\Lambda _{w(0),fl}}} \right)\xi (t).
\end{eqnarray}
Let $\tilde \xi (t) = \left( {\tilde U \otimes {I_{N - 1}}} \right)\xi (t)$, where ${\tilde U^T}\left( {{L_{ff}} + {\Lambda _{w(0),fl}}} \right)U = {\rm{diag}}\left\{ {{{\tilde \lambda }_2},{{\tilde \lambda }_3}, \cdots {{\tilde \lambda }_N}} \right\}$, then it can be found from (\ref{24}) to (\ref{26}) that
\[\dot V(t) \le \sum\limits_{i = 2}^N {\tilde \xi _i^T(t)\left( {RA + {A^T}R - (2{{\tilde \lambda }_i} + {\gamma _l})RB{B^T}R} \right){{\tilde \xi }_i}(t)} .\]
Due to ${L_{ff}} + {\Lambda _{w(0),fl}}$ are positive, one has ${\tilde \lambda _i} > 0{\rm{ }}\left( {i = 2,3, \cdots ,N} \right)$. Hence, if $RA + {A^T}R - {\gamma _l}RB{B^T}R < 0$, then
\[\dot V(t) \le  - \kappa \sum\limits_{i = 2}^N {{{\left\| {{{\tilde \xi }_i}(t)} \right\|}^2}}  =  - \kappa {\left\| {\xi (t)} \right\|^2}\]
for some $\kappa  > 0$. Hence, $\xi (t)$ converges to ${\bf{0}}$ asymptotically; that is, multiagent system (\ref{21}) with protocol (\ref{22}) achieves leader-follower adaptive consensus.\par

In the following, the guaranteed-performance cost is determined. Since the interaction topology among followers is undirected, it can be derived for $h > 0$ that
\[
J_{ff}^h \buildrel \Delta \over = \frac{1}{{N - 1}}\sum\limits_{i = 2}^N {\sum\limits_{k = 2}^N {\int_0^h {{{\left( {{x_k}(t) - {x_i}(t)} \right)}^T}Q\left( {{x_k}(t) - {x_i}(t)} \right){\rm{d}}t} } }
\]\vspace{-1.1em}
\begin{eqnarray}\label{27}
~~~~~~= \int_0^h {{\xi ^T}(t)\left( {2{L_{N - 1}} \otimes Q} \right)\xi (t){\rm{d}}t} .
\end{eqnarray}
Furthermore, due to ${w_{i1}}(0) = 1$, one can show that
\[
J_{fl}^h \buildrel \Delta \over = \sum\limits_{i = 2}^N {\int_0^h {{l_{i1}}{{\left( {{x_1}(t) - {x_i}(t)} \right)}^T}Q\left( {{x_1}(t) - {x_i}(t)} \right){\rm{d}}t} }
\]\vspace{-1.1em}
\begin{eqnarray}\label{28}
{\rm{     }} \hspace{14pt}= \int_0^h {{\xi ^T}(t)\left( {{\Lambda _{w(0),fl}} \otimes Q} \right)\xi (t){\rm{d}}t} .
\end{eqnarray}
Since all the eigenvalues of the Laplacian matrix ${L_{N - 1}}$ of the completed graph with weights $N - 1$ are 1, all the eigenvalues of $2{L_{N - 1}} + {\Lambda _{w(0),fl}}$ are positive and less than 3, so it can be derived by (\ref{27}) and (\ref{28}) that
\begin{eqnarray}\label{29}
J_l^h \buildrel \Delta \over = J_{ff}^h + J_{fl}^h \le \int_0^h {{\xi ^T}(t)\left( {3{I_{N - 1}} \otimes Q} \right)\xi (t){\rm{d}}t} .
\end{eqnarray}
Moreover, it can be shown that
\begin{eqnarray}\label{30}
\int_0^h {\dot V(t)} {\rm{d}}t - V(h) + V(0) = 0.
\end{eqnarray}
Due to ${\lim _{t \to  + \infty }}\left( {{w_{i1}}(t) - {\gamma _{i1}}} \right) = 0$ $\left( {i = 2,3, \cdots ,N} \right)$, one can show that
\begin{eqnarray}\label{31}
\mathop {\lim }\limits_{h \to  + \infty } \sum\limits_{i = 2}^N {\left( {{\gamma _{i1}} - {w_{i1}}(h)} \right)}  = 0.
\end{eqnarray}
If $RA + {A^T}R - {\gamma _l}RB{B^T}R + 3Q \le 0$, then it can be derived from (\ref{29}) to (\ref{31}) that
\begin{eqnarray}\label{32}
\mathop {\lim }\limits_{h \to  + \infty } J_l^h \le {\xi ^T}(0)\left( {{I_{N - 1}} \otimes R} \right)\xi (0) + {\gamma _l}\sum\limits_{i = 2}^N {\left( {{\gamma _{i1}} - {w_{i1}}(0)} \right)} .
\end{eqnarray}
It can be shown that
\begin{eqnarray}\label{33}
{\xi ^T}(0)\left( {{I_{N - 1}} \otimes R} \right)\xi (0) = {x^T}(0)\left( {\left[ {\begin{array}{*{20}{c}}
{N - 1}&{ - {\bf{1}}_{N - 1}^T}\\
{ - {{\bf{1}}_{N - 1}}}&{{I_{N - 1}}}
\end{array}} \right] \otimes R} \right)x(0),
\end{eqnarray}
\begin{eqnarray}\label{34}
\sum\limits_{i = 2}^N {\left( {{\gamma _{i1}} - {w_{i1}}(0)} \right)}  = \sum\limits_{i = 2}^N {\int_0^{ + \infty } {{{\dot w}_{i1}}(t){\rm{d}}t} }  = \int_0^{ + \infty } {{x^T}(t)\left( {\left[ {\begin{array}{*{20}{c}}
{N - 1}&{ - {\bf{1}}_{N - 1}^T}\\
{ - {{\bf{1}}_{N - 1}}}&{{I_{N - 1}}}
\end{array}} \right] \otimes {K_w}} \right)x(t){\rm{d}}t}.
\end{eqnarray}
From (\ref{32}) to (\ref{34}), the conclusion of Theorem \ref{theorem2} can be obtained.$\Box$
\end{pf}

If $(A,B)$ is stabilizable, then $RA + {A^T}R - {\gamma _l}RB{B^T}R + 3Q = 0$ has a unique and positive definite solution. Hence, it is not difficult to find that a necessary and  sufficient condition for leader-follower adaptive guaranteed-performance consensualizability is that $(A,B)$ is stabilizable. Furthermore, ${\gamma _l}$ can be regarded as the rightward translated quantity of the nonzero eigenvalues of $2({L_{ff}} + {\Lambda _{w(0),fl}})$. Moreover, similar to the analysis of Corollary \ref{corollary1}, the following corollary gives an approach to regulate the consensus control gain by introducing a gain factor ${\delta _l} > 0$ such that $R \le {\delta _l}I$.

\begin{col}\label{corollary3}
For any given gain factor ${\delta _l} > 0$, multiagent system (\ref{21}) is leader-follower adaptively guaranteed-performance consensualizable by protocol (\ref{22}) if ${\lambda _{\max }}\left( {B{B^T}} \right) \le 1$ and there exist ${\gamma _l} > 0$ and ${\tilde R^T} = \tilde R \ge \delta _l^{ - 1}I$ such that
\[\tilde \Theta  = \left[ {\begin{array}{*{20}{c}}
{A\tilde R + \tilde R{A^T} - {\gamma _l}B{B^T}}&{3\tilde RQ}\\
*&{ - 3Q}
\end{array}} \right] < 0.\]
In this case, ${K_u} = {B^T}{\tilde R^{ - 1}}$, ${K_w} = {\tilde R^{ - 1}}B{B^T}{\tilde R^{ - 1}}$ and the guaranteed-performance cost satisfies that
\[J_l^* = \sum\limits_{i = 2}^N {\left( {{\delta _l}{{\left\| {{\xi _i}(0)} \right\|}^2} + {\gamma _l}\delta _l^2\int_0^{ + \infty } {{{\left\| {{B^T}{\xi _i}(t)} \right\|}^2}} {\rm{d}}t} \right)} .\]
\end{col}

In the above main results, the Riccati inequality method and the variable changing approach are used to determine gain matrices of consensus protocols. The variable changing approach does not introduce any conservatism because it is an equivalent transformation. However, the Riccati inequality method may bring in some conservatism due to the scalability of the Lyapunov function. It was shown that the Riccati inequality method is extensively used in optimization control and usually has less conservatism in \cite{c33}. Moreover, there are two critical difficulties in obtaining Theorems 1 and 2. The first one is to design a proper Lyapunov function which can translate the nonzero eigenvalues of the Laplacian matrix of the interaction topology rightward. The second one is to construct the relationship between the linear quadratic index and the Laplacian matrix of the interaction topology, as shown in (\ref{13}) and (\ref{14}) for leaderless cases and (\ref{27}) and (\ref{28}) for leader-follower cases.\vspace{0.5em}

\begin{rmk}\label{remark2}
According to the leader-follower structure feature, it is designed that interaction strengths from the leader to followers are adaptively adjusted in protocol (\ref{22}), but interaction strengths among followers are fixed. For leaderless cases, it is required that interaction strengths among all neighboring agents are adaptively time-varying as shown in protocol (\ref{2}). It can be found that fixed interaction weights among followers can simplify the analysis and design of the whole system. Furthermore, it should be pointed that consensus performance among no neighboring agents for both leaderless and leader-follower cases can be adjusted by gain matrices given in Theorems \ref{theorem1} and \ref{theorem2}, which reveals that multiagent systems can regulate global performance by local and indirect interactions. Moreover, compared with the guaranteed-performance cost in \cite{c25,c26,c27,c29}, both $J_r^*$ and $J_l^*$ involve integral terms, which are introduced by adaptively time-varying interaction weights.\vspace{0.5em}
\end{rmk}

\begin{rmk}\label{remark3}
Both the guaranteed-performance cost $J_r^*$ and $J_l^*$ are associated with the initial states of all agents, and the key differences between them are the structures of coupling matrices. It can be found that the coupling matrix ${I_N} - {N^{ - 1}}{{\bf{1}}_N}{\bf{1}}_N^T$ in $J_r^*$ is the Laplacian matrix of a complete graph with edge weights ${N^{ - 1}}$, which means that the guaranteed-performance cost $J_r^*$ is jointly determined by state errors of all agents. However, the coupling matrix $\left[ {\begin{array}{*{20}{c}}
{N - 1}&{ - {\bf{1}}_{N - 1}^T}\\
{ - {{\bf{1}}_{N - 1}}}&{{I_{N - 1}}}
\end{array}} \right]$ in $J_l^*$ is the Laplacian matrix of a star graph with edge weights 1, where the leader is the central node and there do not exist interactions among followers. In this case, the guaranteed-performance cost $J_l^*$ is decided by state errors between the leader and all followers, but it is independent of state errors among followers. Actually, the two coupling matrices intrinsically reflect the impacts of topology structures of multiagent systems on the guaranteed-performance cost.
\end{rmk}

\section{Numerical simulations}\label{section4}
This section presents two numerical simulation examples to demonstrate theoretical results for both leaderless multiagent systems and leader-follower multiagent systems.

\subsection{Numerical simulation for leaderless cases}

\begin{figure}[!htb]
\begin{center}
\scalebox{0.9}[0.9]{\includegraphics{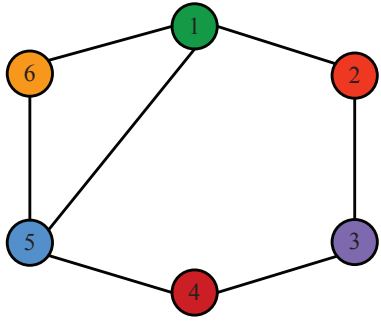}}
\vspace{0em}
\caption{Interaction topology of the leaderless multiagent supporting system.}
\end{center}\vspace{0em}
\end{figure}

Ma {\it et al}. \cite{c34} proposed the model of the multiagent supporting system which has potential applications in earthquake damage-preventing buildings, water-floating plants and large-diameter parabolic antennae, and gave a centralized control approach. Xi et {\it et al}. \cite{c35} presented a distributed approach to control the multiagent supporting system, where the dynamics of each agent with the self feedback matrix $\left[ { - 88.6792,4.7642} \right]$ can be modeled as (\ref{1}) with
\[
A = \left[ {\begin{array}{*{20}{c}}
   0 & 1  \\
   { - 100} & 0  \\
\end{array}} \right],
B = \left[ \begin{array}{l}
 0 \\
 1 \\
 \end{array} \right],
\]
and the interaction topology of this leaderless multiagent supporting system is shown in Fig. 1 with all edge weights equal to 1. Let
\[
Q = \left[ {\begin{array}{*{20}{c}}
   1 & 0  \\
   0 & 2  \\
\end{array}} \right],
\]
then one can obtain from Theorem 1 that
\[
P = \left[ {\begin{array}{*{20}{c}}
   {{\rm{223}}{\rm{.5978}}} & {{\rm{3}}{\rm{.9324}}}  \\
   {{\rm{3}}{\rm{.9324}}} & {{\rm{2}}{\rm{.1307}}}  \\
\end{array}} \right],
\]
\[
{K_u} = \left[ {{\rm{3}}{\rm{.9324}},{\rm{2}}{\rm{.1307}}} \right],
\]
\[
{K_w} = \left[ {\begin{array}{*{20}{c}}
   {{\rm{15}}{\rm{.4638}}} & {{\rm{8}}{\rm{.3788}}}  \\
   {{\rm{8}}{\rm{.3788}}} & {{\rm{4}}{\rm{.5399}}}  \\
\end{array}} \right],
\]
and the guaranteed-performance cost is $J_r^* = 1694.6$.
\vspace{1.0em}
\begin{figure}[!htb]
\begin{center}
\scalebox{0.45}[0.45]{\includegraphics{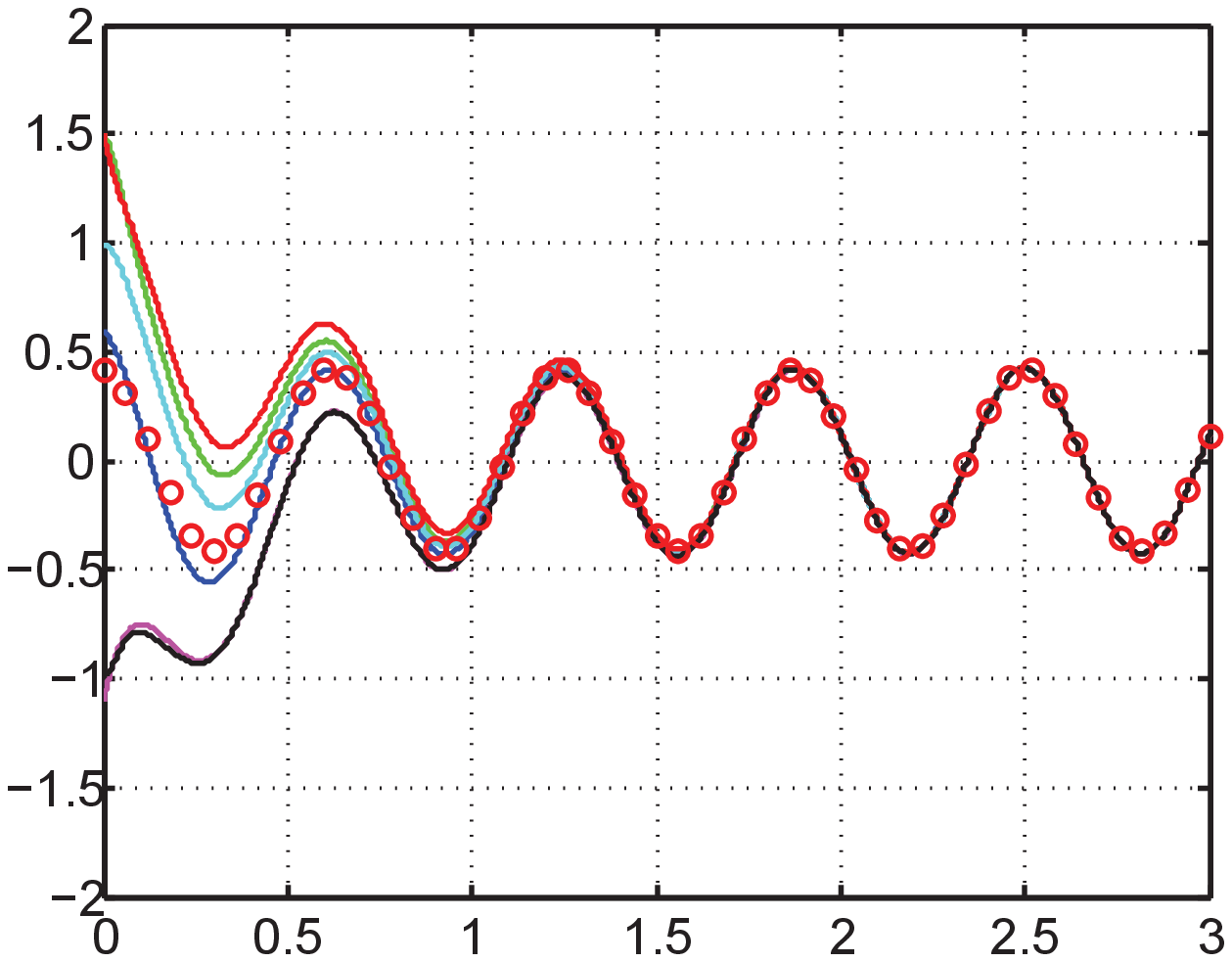}}
\scalebox{0.45}[0.45]{\includegraphics{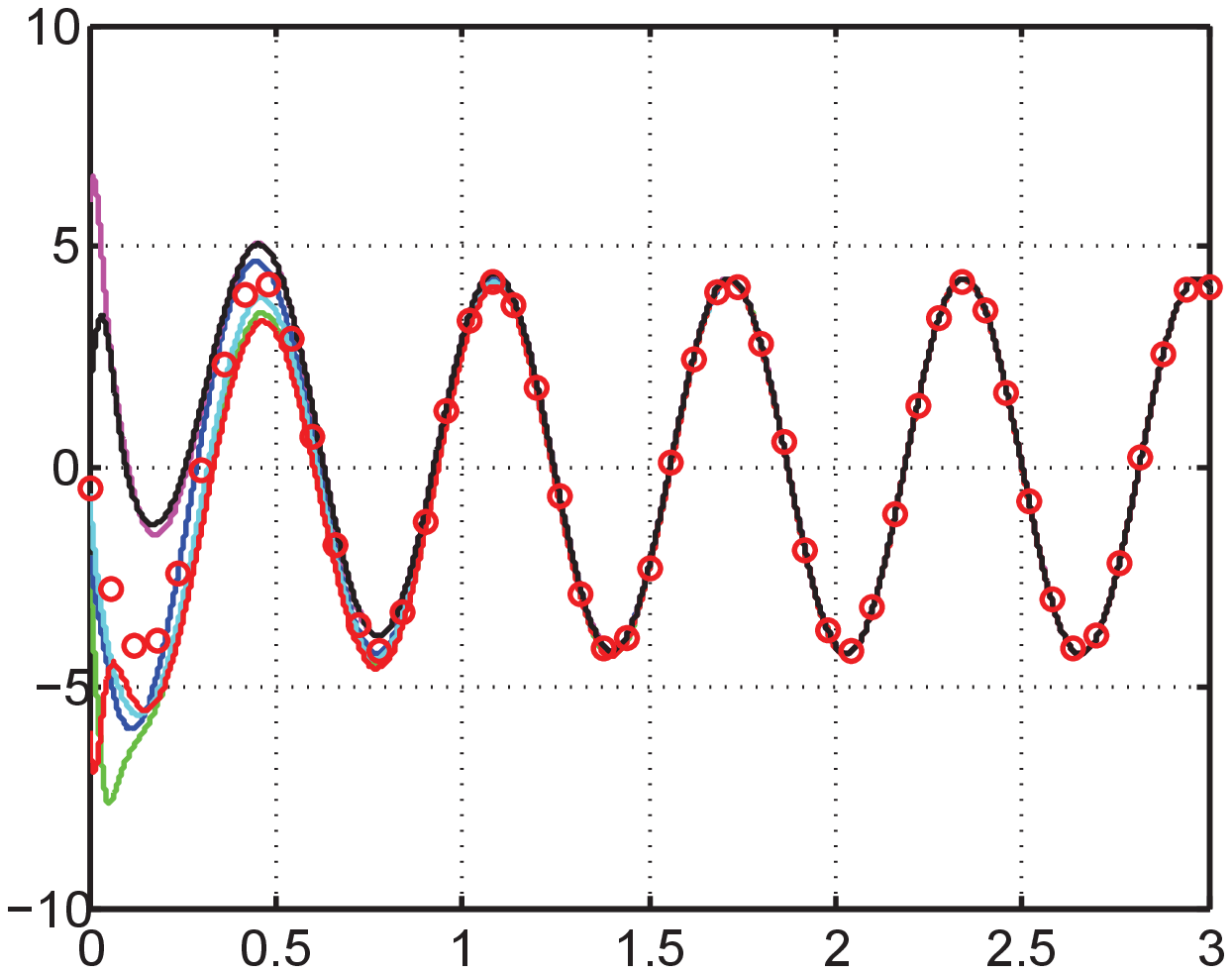}}
\put (-370, 60)
{\rotatebox{90} {{\scriptsize ${x_1}(t)$}}} \put (-280, -5) {{ \scriptsize {\it t}~/~\it s}}
\put (-185, 60)
{\rotatebox{90} {{\scriptsize ${x_2}(t)$}}} \put (-97, -5) {{ \scriptsize {\it t}~/~\it s}}
\vspace{0em} \caption{State trajectories of the leaderless multiagent supporting system.}
\end{center}\vspace{-0.5em}
\end{figure}

\begin{figure}[!htb]
\begin{center}
\scalebox{0.5}[0.5]{\includegraphics{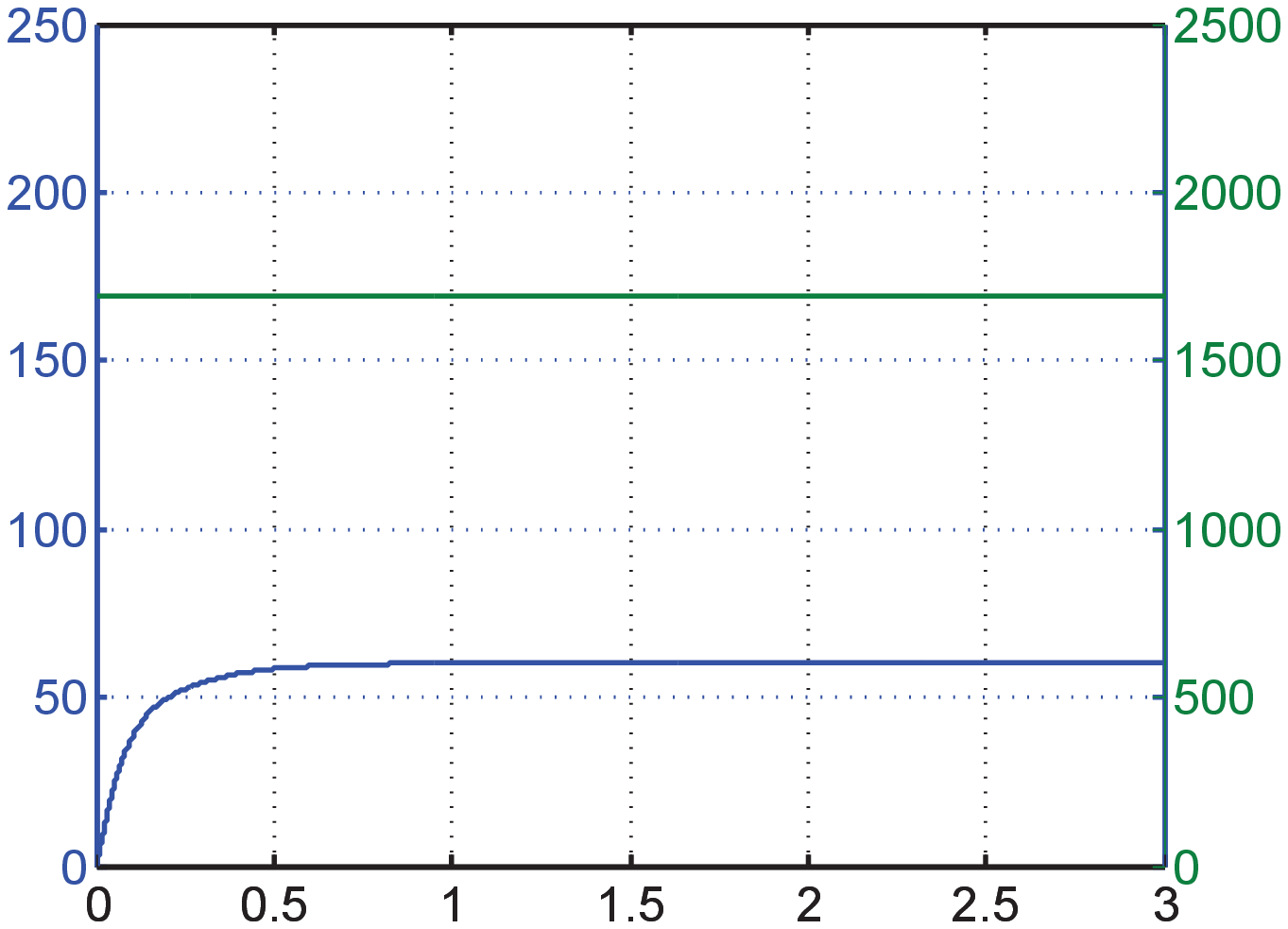}} 
\put (-105, -5){{\scriptsize {\it t}~/~\it s}}
\put (-105, 62) {{ \scriptsize ${J_r}$}}
\put (-105, 110) {{ \scriptsize $J_r^*$}}
 \vspace{0em}
\caption{Performance function of the leaderless multiagent supporting system.}
\end{center}\vspace{-0em}
\end{figure}

Fig. 2 depicts the state trajectories of the leaderless multiagent supporting system, where the trajectories marked by circles are the curves of the consensus function given in Corollary 2 and the trajectories of states of six agents are represented by full curves. The trajectory of the performance function  $J_r^h$ with $h=3$ is shown in Fig. 3. One can see that state trajectories of all agents converge to the ones marked by circles and the performance function $J_r^h$ converges to a finite value with ${J_r^h} < J_r^*$; that is, this multiagent supporting system achieves leaderless adaptive guaranteed-performance consensus without using the global information of the interaction topology. However, the distributed approach in \cite{c35} required the precise value of the minimum eigenvalue of the interaction topology; that is, the completely distributed control cannot be realized.

\subsection{Numerical simulation for leader-follower cases}
\begin{figure}[!htb]
\begin{center}
\scalebox{0.9}[0.9]{\includegraphics{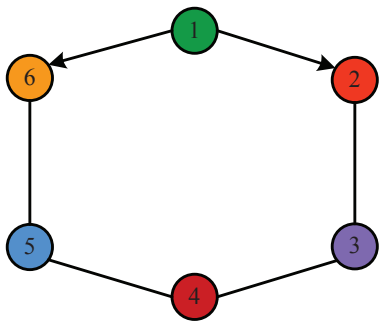}}
\vspace{0em}
\caption{Interaction topology of the leader-follower multiagent system.}
\end{center}\vspace{-1em}
\end{figure}
Consider a leader-follower multiagent system with one leader and five followers, where the interaction topology is shown in Fig. 4 with all edge weights equal to 1 and the dynamics of each agent is modeled by (\ref{21}) with
\[
A = \left[ {\begin{array}{*{20}{c}}
   1 & 1 & 0 & 0  \\
   { - 30} & { - 12.5} & {30} & 0  \\
   0 & {0.5} & 0 & 1  \\
   {16} & 0 & { - 16} & 0  \\
\end{array}} \right],
~B = \left[ {\begin{array}{*{20}{c}}
   1  \\
   {19}  \\
   0  \\
   0  \\
\end{array}} \right].
\]

Let
\[
Q = \left[ {\begin{array}{*{20}{c}}
   {0.30} & {0.30} & {0.20} & {0.10}  \\
   {0.30} & {0.50} & {0.10} & {0.10}  \\
   {0.20} & {0.10} & {0.50} & {0.15}  \\
   {0.10} & {0.10} & {0.15} & {0.10}  \\
\end{array}} \right],
\]
then one can obtain by Theorem \ref{theorem2} that
\[
R = \left[ {\begin{array}{*{20}{c}}
   {{\rm{7}}{\rm{.7420}}} & { - {\rm{0}}{\rm{.1280}}} & { - {\rm{6}}{\rm{.0953}}} & {{\rm{0}}{\rm{.6304}}}  \\
   { - {\rm{0}}{\rm{.1280}}} & {{\rm{0}}{\rm{.0404}}} & {{\rm{0}}{\rm{.1680}}} & {{\rm{0}}{\rm{.0148}}}  \\
   { - {\rm{6}}{\rm{.0953}}} & {{\rm{0}}{\rm{.1680}}} & {{\rm{7}}{\rm{.0299}}} & {{\rm{0}}{\rm{.1259}}}  \\
   {{\rm{0}}{\rm{.6304}}} & {{\rm{0}}{\rm{.0148}}} & {{\rm{0}}{\rm{.1259}}} & {{\rm{0}}{\rm{.7516}}}  \\
\end{array}} \right],
\]
\[
{K_u} = \left[ {{\rm{5}}{\rm{.3100}},{\rm{ 0}}{\rm{.6396}},{\rm{ }} - {\rm{2}}{\rm{.9033}},{\rm{ 0}}{\rm{.9116}}} \right],
\]
\[
{K_w} = \left[ {\begin{array}{*{20}{c}}
   {{\rm{28}}{\rm{.1961}}} & {{\rm{3}}{\rm{.3963}}} & { - {\rm{15}}{\rm{.4165}}} & {{\rm{4}}{\rm{.8406}}}  \\
   {{\rm{3}}{\rm{.3963}}} & {{\rm{0}}{\rm{.4091}}} & { - {\rm{1}}{\rm{.8570}}} & {{\rm{0}}{\rm{.5831}}}  \\
   { - {\rm{15}}{\rm{.4165}}} & { - {\rm{1}}{\rm{.8570}}} & {{\rm{8}}{\rm{.4292}}} & { - {\rm{2}}{\rm{.6466}}}  \\
   {{\rm{4}}{\rm{.8406}}} & {{\rm{0}}{\rm{.5831}}} & { - {\rm{2}}{\rm{.6466}}} & {{\rm{0}}{\rm{.8310}}}  \\
\end{array}} \right],
\]
and the guaranteed-performance cost is $J_l^* = {\rm{872}}{\rm{.5}}.$\par

\begin{figure}[!htb]
\begin{center}
\scalebox{0.45}[0.45]{\includegraphics{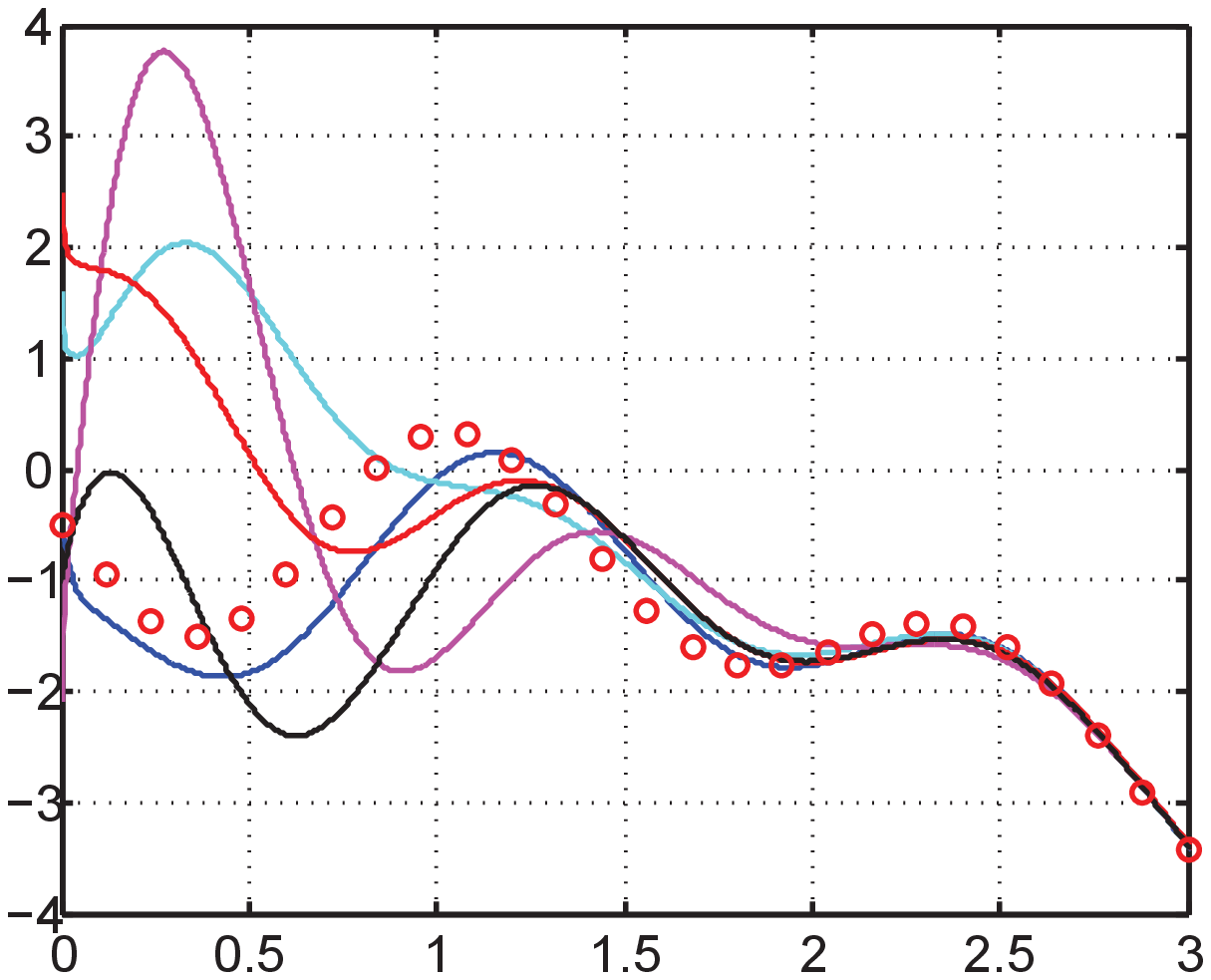}}
\scalebox{0.45}[0.45]{\includegraphics{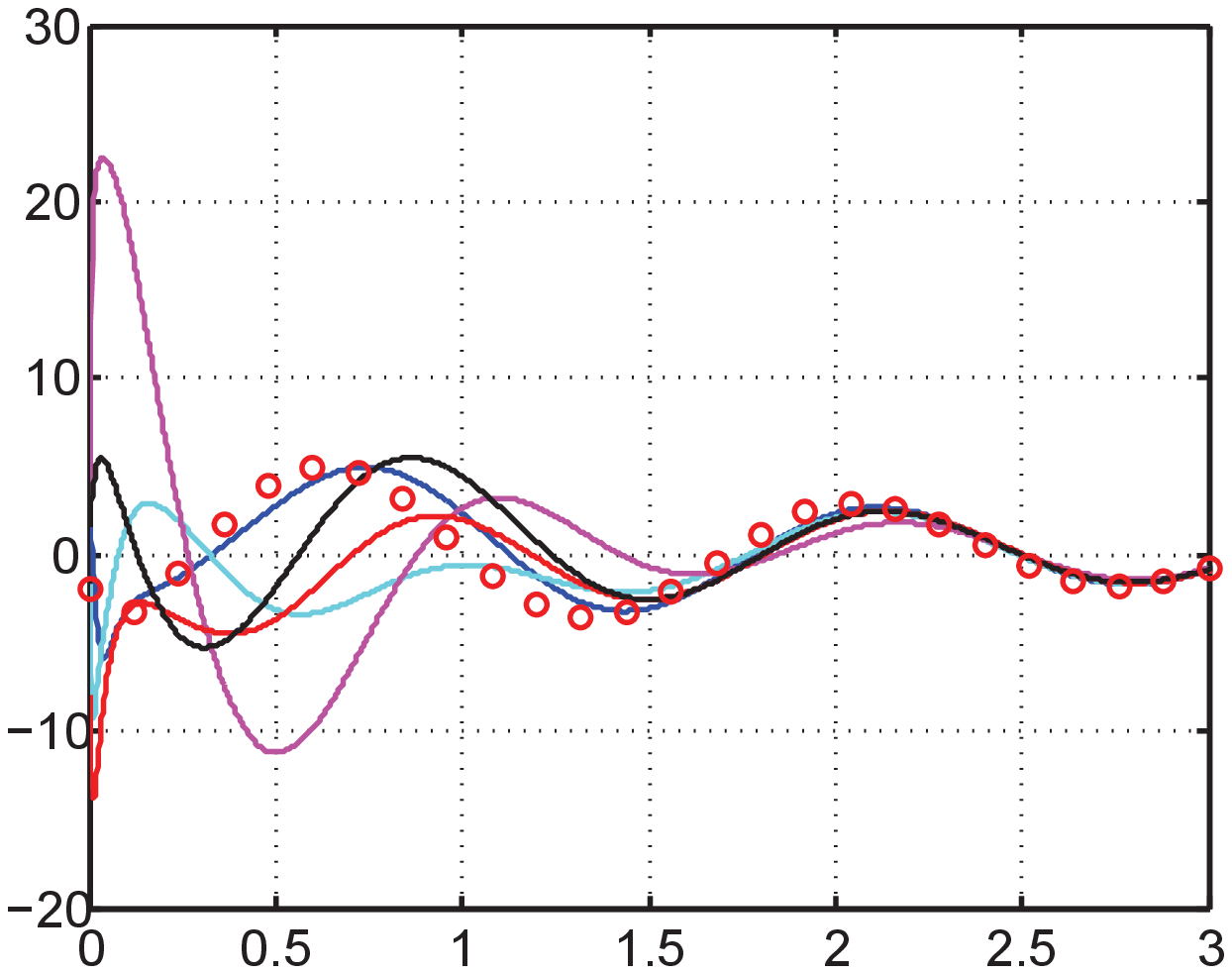}}
\put (-370, 60)
{\rotatebox{90} {{\scriptsize ${x_1}(t)$}}} \put (-280, -5) {{ \scriptsize {\it t}~/~\it s}}
\put (-185, 60)
{\rotatebox{90} {{\scriptsize ${x_2}(t)$}}} \put (-97, -5) {{ \scriptsize {\it t}~/~\it s}}
\end{center}\vspace{-2.5em}
\end{figure}

\begin{figure}[!htb]
\begin{center}
\scalebox{0.45}[0.45]{\includegraphics{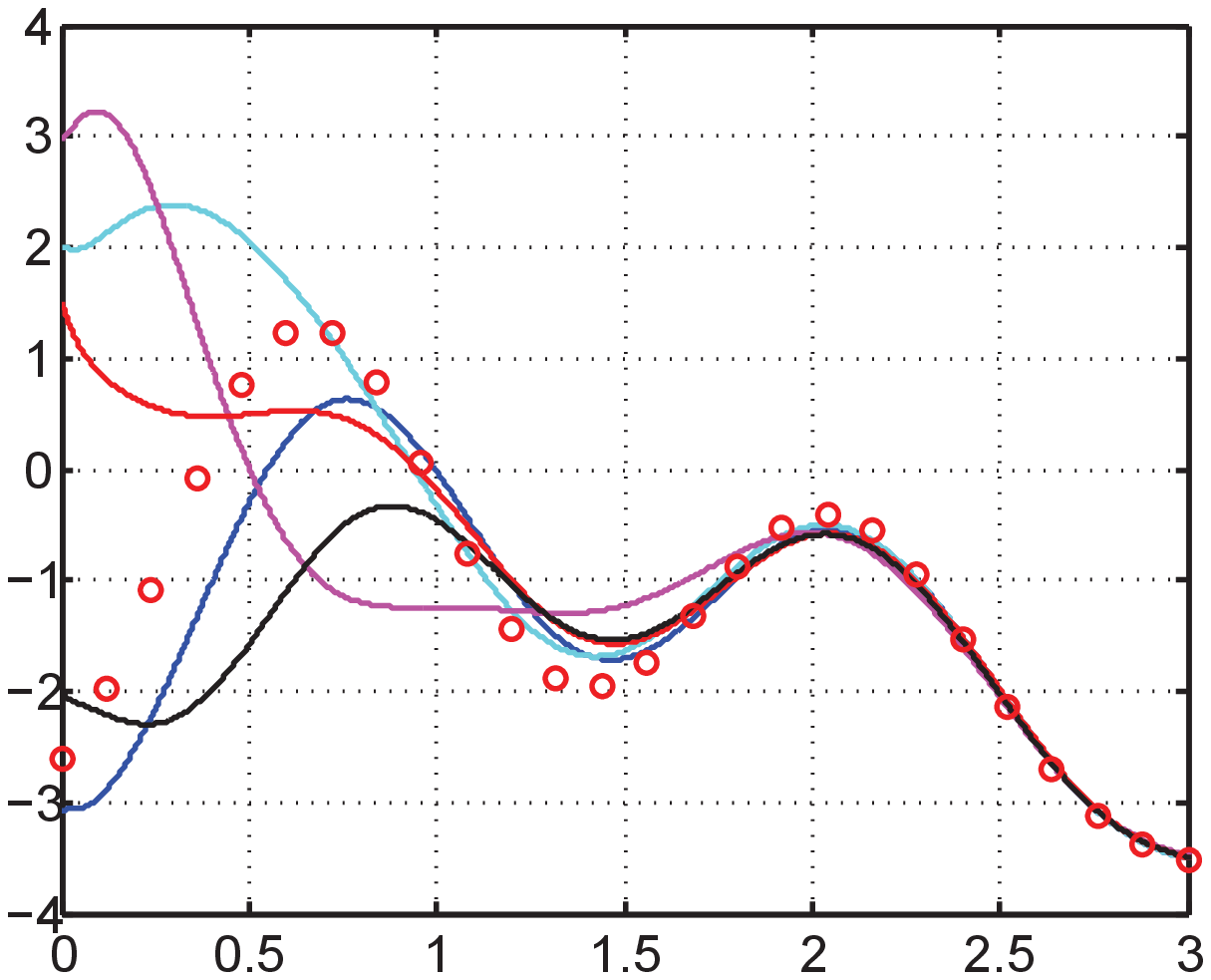}}
\scalebox{0.45}[0.45]{\includegraphics{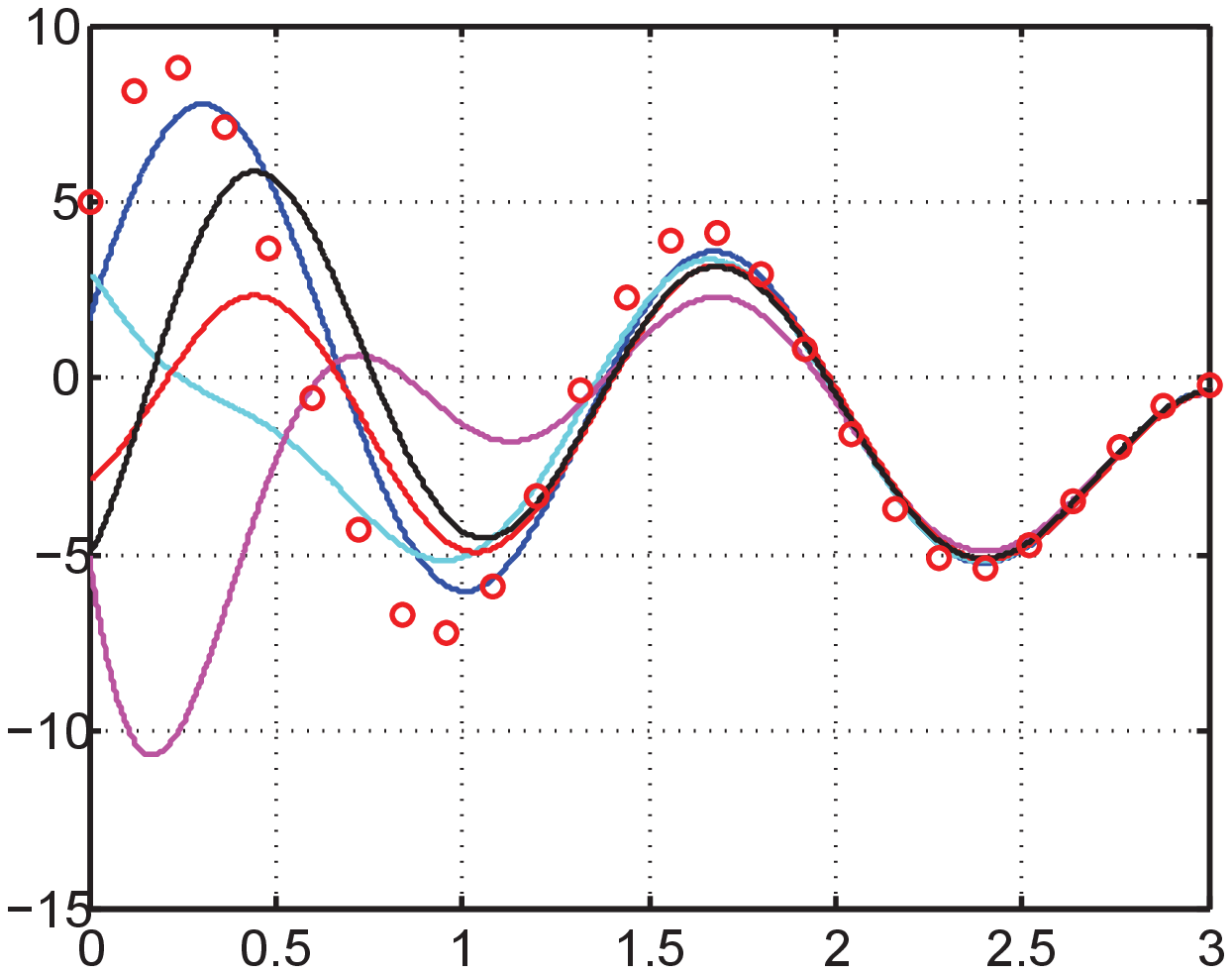}}
\put (-370, 60)
{\rotatebox{90} {{\scriptsize ${x_3}(t)$}}} \put (-280, -5) {{ \scriptsize {\it t}~/~\it s}}
\put (-185, 60)
{\rotatebox{90} {{\scriptsize ${x_4}(t)$}}} \put (-97, -5) {{ \scriptsize {\it t}~/~\it s}}
\vspace{0em} \caption{State trajectories of the leader-follower multiagent system.}
\end{center}\vspace{-0.5em}
\end{figure}

\begin{figure}[!htb]
\begin{center}
\scalebox{0.5}[0.5]{\includegraphics{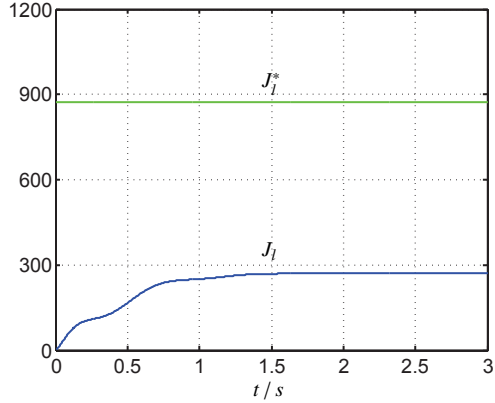}} 
\put (-105, -5){{\scriptsize {\it t}~/~\it s}}
\put (-105, 48) {{ \scriptsize ${J_l}$}}
\put (-105, 112) {{ \scriptsize $J_l^*$}}
 \vspace{0em}
\caption{Performance function of the leader-follower multiagent system.}
\end{center}\vspace{-1em}
\end{figure}

The state trajectories of the leader-follower multiagent system are shown in Fig. 5, where the trajectories marked by circles denote the states of the leader and the trajectories of states of five followers are represented by full curves.  In Fig. 6, the trajectory of the performance function $J_l^h$ with $h=3$ is shown. It can be found that state trajectories of all followers converge to the ones of the leader and the performance function $J_l^h$ converges to a finite value with ${J_l^h} < J_l^*$, which means that this multiagent system achieves leader-follower adaptive guaranteed-performance consensus without using the global information of the interaction topology. Moreover, it should be pointed out that the computational complexity does not increase as the number of agents increases since the main results of the current paper are completely distributed.

\section{Conclusions}\label{section5}
A new adaptive consensus scheme with a quadratic cost function was proposed to realize the completely distributed guaranteed-performance consensus control. The adaptive guaranteed-performance consensualization criterion for the leaderless case was given by regulating the interaction weights among all neighboring agents, and the adaptive guaranteed-performance consensualization criterion for the leader-follower case was presented by regulating the interaction weights from the leader to its followers. Furthermore, for leaderless and leader-follower multiagent systems, the regulation approaches of the consensus control gain were proposed, respectively, which adjust control gains by choosing the different translation factors. Moreover, explicit expressions of the guaranteed-performance costs were given, where the coupling matrix associated with the leaderless case is the Laplacian matrix of a complete graph and the coupling matrix associated with the leader-follower case is the Laplacian matrix of a star graph with the leader being the central node.\par

Furthermore, the future research directions can focus on two aspects. The first one is to study the practical applications of multiagent systems combining main results in the current paper with structure characteristics of practical multiagent systems, such as network congestion control systems and single-link manipulator systems with a flexible joint, {\it et al}. The other one is to investigate the impacts of directed topologies, time-varying delays and given cost budgets on adaptive guaranteed-performance consensus of multiagent systems.
\vspace{-0.5em}

\bibliographystyle{elsarticle-num}


%

\newpage

\end{document}